\documentclass{emulateapj}
\usepackage{color}
\usepackage{latexsym, graphicx, amssymb, longtable,epsf,ulem}
\usepackage{url}
\usepackage{color}
\newcommand{\xxx}{\color{blue}}
\newcommand{\xx}{\color{black}}
\newcommand{\x}{\sout}

\newcommand{\w}{\color{black}}
\newcommand{\xie}{\color{black}}
\newcommand{\gr}{\color{black}} 
\newcommand{\wyq}{\color{black}} 

\long\def\symbolfootnote[#1]#2{\begingroup\def\thefootnote{\fnsymbol{footnote}}\footnote[#1]{#2}\endgroup}

\shorttitle{TTV and Multiplicity of KOI}
\shortauthors{Xie et al.}

\begin{document}

\title{Frequency of Close Companions among Kepler Planets -- a TTV study
}

\author{Ji-Wei Xie$^{1, 2}$, Yanqin Wu$^1$, Yoram
  Lithwick$^{3}$}
\affil{$^1$Department of
  Astronomy and Astrophysics, University of Toronto, Toronto, ON M5S
  3H4, Canada; \\ jwxie@astro.utoronto.ca; wu@astro.utoronto.ca}
\affil{$^2$Department of Astronomy \& Key Laboratory of Modern
  Astronomy and Astrophysics in Ministry of Education, Nanjing
  University, 210093, China} \affil{$^3$Department of Physics and
  Astronomy, Northwestern University, 2145 Sheridan Rd., Evanston, IL
  60208 \& Center for Interdisci- plinary Exploration and Research in
  Astrophysics (CIERA); y-lithwick@northwestern.edu}



\begin{abstract}
  A transiting planet exhibits sinusoidal transit-time-variations
  (TTVs) if perturbed by a companion near a mean-motion-resonance
  (MMR).  
We search for sinusoidal TTVs in more than 2600 Kepler candidates,
using the publicly available Kepler light-curves (Q0-Q12). We find
that the TTV fractions rise strikingly with the transit
multiplicity. Systems where four or more planets transit enjoy
\x{four} {\w roughly five} times higher TTV fraction than those where
a single planet transits, and about twice higher than those for
doubles and triples. In contrast, models in which all transiting
planets arise from similar dynamical configurations predict comparable
TTV fractions among these different systems. 
{\w One simple explanation for our results is} that there
are at least two different classes of Kepler systems, one closely
packed and one more sparsely populated.
\end{abstract}



\section{Introduction}\label{SEC:INTRO}

Since the launch in March 2009, the {\it Kepler} mission has
discovered a few thousand planetary candidates, called Kepler Objects
of Interests (KOIs), by detecting the flux deficit as a planet
transits in front of its star \citep{Bor11, Bat13, OD13, Hua13,
  Bur13}. While some of the stars are observed to
  have one transiting planet \citep[called ``tranet''
from now on, following][]{TD12}, others show up to 6
\citep{Lis11a}.  A natural question to ask is, do all of these systems
share the same intrinsic orbital structures?  For observing
transiting planets, the two most relevant orbital parameters are the
dispersion in orbital inclinations, and the typical spacing between
adjacent planets.

\begin{figure*}
\begin{center}
\includegraphics[width=\textwidth]{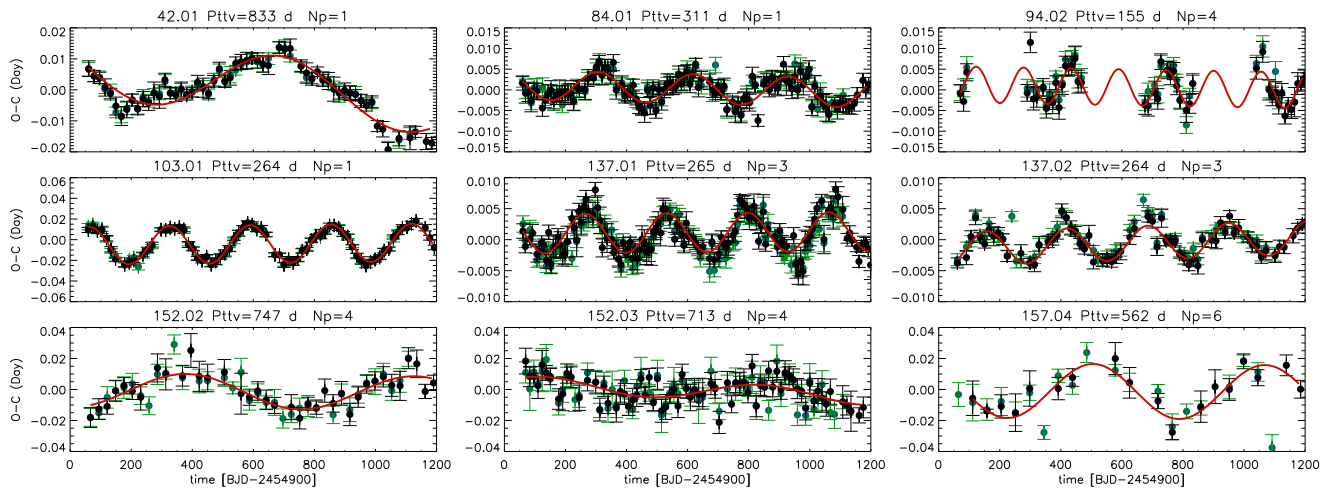}
\caption{A subset of TTV candidates identified with our standard
  criterion in the reduced KOI sample (case 1). The title of each
  panel indicates the KOI number, TTV super-period in days, and number
  of tranets in that system.  In each panel we plot our measured TTV
  data points (black data) on top of the TTV measurements from
  \citet{Maz13} (green data if any) as well as the best sinusoidal fit
  (red line).  For data on all of our TTV candidates, see
  http://www.astro.utoronto.ca/$\sim$jwxie/TTV.  }
\label{fig:ttv}
   \end{center}
\end{figure*}

A number of groups have studied the inclination dispersion of Kepler
planets and reached the common conclusion that this must be small and
is of order a few degrees {\xie \citep{Lis11b, FM12,TD12, Fab12a,
    Fig12, Joh12}}. However, it has been pointed out that models with
a single inclination dispersion falls short in explaining the number
of single tranets relative to higher multiples, by a factor of three
or more \citep{Lis11b}. This suggests that all Kepler planets are not
the same, and motivates models where the inclination dispersion itself
is broadly distributed \citep[``Rayleigh of
Rayleigh'',][]{Lis11b,Fab12a}.  However, the relative occurrences of
different Kepler multiples (denoted here as 1P, 2P, 3P... by the
number of tranets seen in a system) are sensitive to both the
inclination dispersion and the intrinsic planet spacing.  Larger
spacing between adjacent planets will raise the relative number of
single tranet systems, {\w as} \x{so} will larger inclination
dispersion.  It is difficult to disentangle the two without the aid of
further information.  Therefore we turn to a new measure, the TTV
fraction.

  If a tranet is accompanied by another planet, its transit times
  deviate from strict periodicity \citep[transit-time-variation,
  TTV,][]{HM05,Ago05}. Many studies have used TTV to confirm the
  planetary nature of Kepler candidates \citep[e.g.][] {Hol10, Lis11a,
    Coc11, Bal11, For12a, Ste12, Fab12b, Car12, Nes12, Xie13,Xie14,
    Ste13}. Furthermore, it is realized that when the companion is
  near a mean-motion resonance (MMR) with the tranet, the TTV is
  particularly strong and exhibits a characteristic sinusoidal form
  \citep{Ago05}. The amplitude and phase of this sinusoid 
    have been simply related to the perturber's mass, as well as the
  orbital eccentricities \citep{Lit12}, thereby allowing us to infer
  the interior composition and orbital parameters of these objects
  {\w \citep{WL13, HL13}}.

  Just as the TTV signal can be used to infer the presence of unseen
  (non-transiting) companions around specific candidates
  \citep[e.g.][]{Nes12,Nes13}, the 
  {\w number} of tranets that exhibit sinusoidal TTVs 
  {\xie provides constraints on} near-MMR
  companions.  Since the period ratios of adjacent Kepler pairs do not
  much prefer MMRs \citep{Fab12a}, these near-MMR companions can be
  taken as a proxy for companions that lie close to and inward of the 2:1
  MMR.
  
   To be quantitative, we shall define the ``intrinsic TTV
    fraction'' as half the probability that a planet induces a
    sufficiently large TTV amplitude {\xx for detection} \footnote{ {\w The TTV
      amplitudes for a pair of planets are determined by physical
      properties such as masses, eccentricities, and the distance to
      resonance.}}  {\gr in another planet in the system.  The reason
      for the factor of a half is that when one planet has a large
      TTV, then typically so does its TTV partner, and we do not wish
      to double-count such a pair.  When trying to measure this
        quantity observationally, we shall first count the number of
      observed tranets with measured TTV's, and then subtract one each
      time two tranets are TTV partners.  Dividing by the total number
      of tranets yields the ``measured TTV fraction,'' which is our
      estimate for the intrinsic fraction. }

      {\w Our ability to measure TTV is affected by the noise level in the
      transit signals, which is in turn determined by a range of
      parameters including stellar brightness, the size of the planet
      relative to its host star, the orbital period and the transit
      duration. }  
     {\w However, if we split the planet candidates into
      different groups, and if these groups share the same noise
      properties, then one can argue that the relative TTV fractions
      measured for different groups represent the relative differences
      in their intrinsic TTV fractions.}

    In the following, we proceed to measure the relative TTV
    fraction{\gr s} among 1P, 2P, 3P and 4P+ systems, where 4P+ stands
    for systems that have four or more transiting planets. We carry
    out the analysis for all KOIs that have suitable light-curves,
    which include more than 2600 KOIs. We interpret the significance
    of our results in \S \ref{sec:discussion}.

\begin{figure*}
\begin{center}
\includegraphics[width=0.8\textwidth]{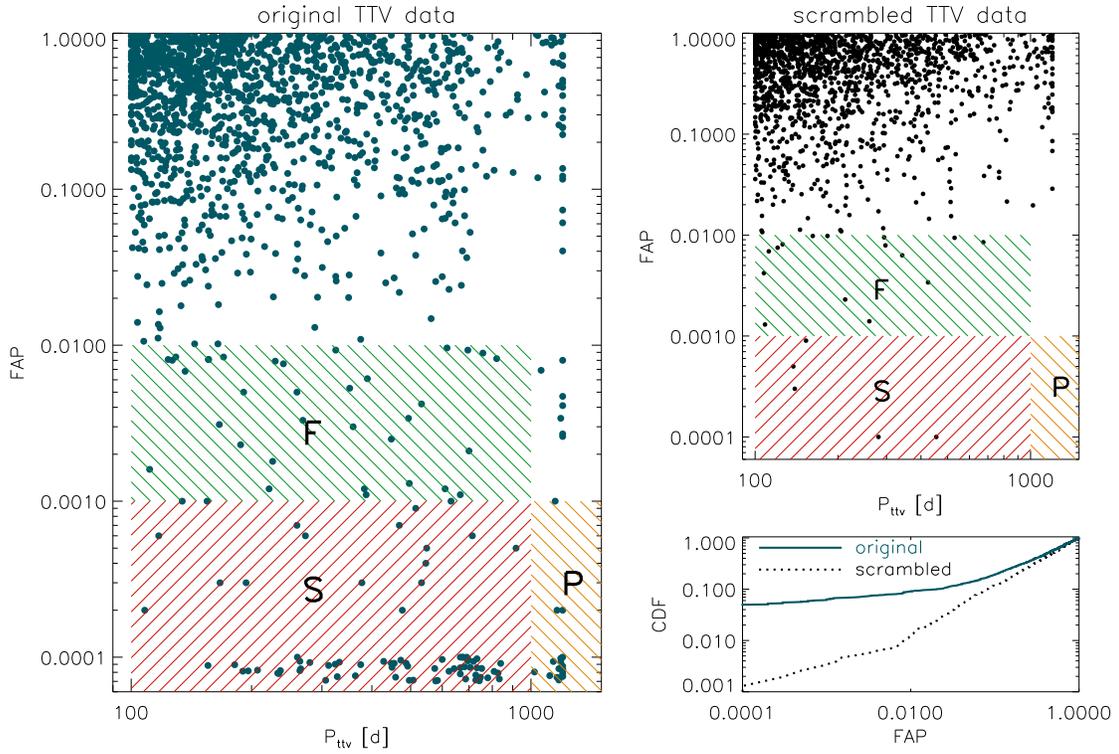}
\caption{The TTV periods and FAPs obtained for the full sample of 2606
  KOIs.  The left panel is for the original TTV data, while the right
  top panel pertains to a `scrambled' population, one which has the same TTV data as the
    observed ones but with randomly scrambled time stamps. This set
    is equivalent to pure noise and acts as a control sample.  For the sake of clarity, all points with
  FAP$<10^{-4}$ are displayed just below that value, with a slight 
    artificial dispersion.  The red
  hatched region labelled as `S' illustrates our conservative
  criterion for identifying TTV candidates (\S \ref{subsec:identify}),
  while more relaxed criteria include regions `F' or `P'.  The
  right-bottom panel shows the cumulative distribution of FAPs for the
  original TTV data (solid) and for the scrambled data (dashed). The
  latter satisfies CDF = FAP, as should be the case for pure noise.
 The real data show an excess of small FAP objects, over that of
    noise, corresponding to genuine TTV candidates. Small FAP cases,
    for the scrambled data, occur preferentially at short TTV periods.
}
\label{fig:pttv}
   \end{center}
\end{figure*}

\section{Measuring TTV Fractions}
\label{sec:method}

\subsection{Transit time measurements}

We use the publicly available Q0-Q12 long cadence (LC, PDC) data for
2740 KOIs \citep[Kepler objects of interest,][]{Bur13}. 
Out of these, 134 KOIs have fewer than 7 transit time measurements,
either because the transit periods are very long, or the signal-to-noise ratios (SNR) are too
small. These spread evenly across all multiplicities. This leaves us
with 2606 KOIs, out of which there are, 1488, 571, 320, 227  systems that
are designated as 1P, 2P, 3P and 4P+, respectively. We refer to this
sample of 2606 KOIs as the `full' sample. 

We have also selected a `reduced' sample by excluding those KOIs for
which timing measurements are less accurate.  These include those with
large noise ($SNR \leq 15$), {\xie and} short transit duration (less
than an hour).
We include only KOIs with intermediate planet sizes ($0.8 R_\oplus \leq
R_p \leq 8 R_\oplus$) as they likely have the lowest false positive
rate \citep{Fre13}.  This reduced sample contains a total of 
{\xie 1989 KOIs, with 1097, 446,
  253, and 193} systems designated as 1P, 2P, 3P, and 4P+,
respectively.

The pipeline used to measure the transit times has been developed and
described in \citet{Xie13}. We compare our transit time measurements
to the published ones \citep{For12a, For12b, Ste12, Fab12a, Maz13},
and found good consistency (see, e.g., Fig.\ref{fig:ttv}).  Our
  measurements for the TTV candidates are publicly available at
  http://www.astro.utoronto.ca/$\sim$jwxie/TTV.

\subsection{Identification of  sinusoidal TTV}
\label{subsec:identify}
From the above transit time measurements, we derive TTV, which are the
residuals after a best linear fit.  We then search for a sinusoidal
signal by obtaining a Lomb-Scargle (LS) periodogram \citep{Sca82,
  ZK09} on these residuals and identify the highest peak that has a
period longer than twice the orbital period, as well as $>100$
days. The former threshold comes about because twice the orbital
period is the Nyquist frequency for sampling TTV. The latter threshold
is enforced because TTV at shorter periods can be significantly
polluted by noise from chromospheric activities, as stellar rotation
periods fall typically in the range from a few to a few tens of days
\citep{Sza13,Maz13}.

Sinusoidal TTV caused by a perturber near a MMR has a ``super-period''
{\xie \citep{Ago05, Lit12}},
\begin{equation}
P_{\rm ttv} \equiv {1\over{|j/P' - (j-1)/P|}} = {{P'}\over{j |\Delta|}},
\label{eq:superp}
\end{equation}
where $P$ and $P'$ are the orbital periods of the two planets that are
near a first-order $(j:j-1)$ MMR
by a fractional distance $\Delta$. For a planet pair with period
$P'=10$ days, $j=2$, $|\Delta| \sim 5\%$, we find $P_{\rm ttv} \sim
100$ days.  A planet pair with a larger $|\Delta|$ will have a shorter
super-period, however their TTVs also become increasingly difficult
to detect as the TTV amplitudes scale inversely with $|\Delta|$
{\xie \citep{Ago05, Lit12}}. All reported cases
\citep{Lit12,Xie13,Xie14,WL13,Ste12,Rag12} have $|\Delta|$
falling between $ 1 - 5\%$.  This consideration,
coupled with the above concern for chromospheric noise, leads us to
discard sinusoids short-ward of 100 days.  

We also adopt an upper limit of $P_{\rm ttv} \leq 1000$ days. This
comes about because the data (Q0-Q12) stretch only $\sim 1100$
days. However, some TTV systems show strong, identifiable sinusoids
even before a full TTV cycle is observed.  This constraint is later
relaxed and is found not to impact the conclusion.

Many of the sinusoids thus identified are false, caused by random
alignment of noisy data. It is an important task to exclude these.  We
adopt the following strategy from \citet{Cumming2004}, originally
applied to detect planets from radial velocity data. For each KOI, we
scramble the time stamps of the original TTV data for $10^4$
times, perform a LS periodogram analysis on each set of data, obtaining
the amplitude and frequency on the highest peak. The FAP (false alarm
probability) of the original TTV peak is estimated as the fraction of
permutations that have higher sinusoids than the original TTV.  We
assign an FAP value of $10^{-4}$ if not a single random realization
exceeds the observed sinusoid amplitude.  Our FAP estimates compare
well with those from \citet{Maz13}.

Fig.\ref{fig:pttv} shows the FAP and $P_{\rm ttv}$ for each of the
KOIs in our full sample,  as well as those using a scrambled
  time series for the same KOIs. The latter set is equivalent to
  random noise and so acts as a control sample.  The true data
show a significant excess of objects at very low FAPs, when compared
to those of the scrambled data.  We adopt the following `standard'
criterion (the region labelled as `S' in Fig. \ref{fig:pttv}) for
identifying our TTV candidates:
\begin{itemize}
\item (1) $\rm FAP <10^{-3}$ and
\item (2) TTV period between 100 and 1000 d.
\end{itemize}
Objects that have FAP $\leq 10^{-3}$ exhibit TTV amplitudes that range
from one to hundreds of minutes, with TTV sensitivity higher for larger
SNR objects.  Our above FAP criterion is on the conservative side: for
an FAP of $10^{-3}$, there should only be $2606\times 10^{-3} \sim 3$
false positives among our TTV candidates, much fewer than the actual
number of candidates ($\sim 100$). We experiment by relaxing the above
criterion, either by raising the FAP threshold to $10^{-2}$ (adding
the `F' region in Fig. \ref{fig:pttv}), or by removing the 1000 day
upper limit (adding the `P' region). We report our results below.

\begin{table*}[]
 \begin{center}
  \caption{TTV fraction as a function of transit multiplicity}
\begin{tabular}{|c|cc|cc|cc|cc|cc|}
\hline
case & KOIs$^{a}$ & TTVs$^{a}$ & \multicolumn{2}{|c|}{1P}  & 
\multicolumn{2}{|c|}{2P}   & \multicolumn{2}{|c|}{3P}  & \multicolumn{2}{|c|}{4P+} \\ 
 &  &  & 
$N_{\rm ttv}/N_{\rm koi}$$^{b}$  & $\%$$^{c}$  & 
$N_{\rm ttv}/N_{\rm koi}$  & $\%$  & 
$N_{\rm ttv}/N_{\rm koi}$  & $\%$  & 
$N_{\rm ttv}/N_{\rm koi}$  & $\%$  \\
\hline
0 & full & S & (31)31/1488  & 2.1$\pm$0.4 & (24)19/571  & 3.3$\pm$0.8 & (19)13/320 & 4.1$\pm$1.1 & (23)17/227  & 7.5$\pm$1.8  \\
1 & reduced & S & (19)19/1097  & 1.7$\pm$0.4 & (24)19/446 & 4.3$\pm$1.0 & (16)13/253  & 5.1$\pm$1.4 & (23)17/193 & 8.8$\pm$2.1 \\
2 & reduced & S + P & (26)26/1097 & 2.4$\pm$0.5 & (32)26/446 & 5.8$\pm$1.1 & (22)16/253 & 6.3$\pm$1.6 & (28)21/193 & 10.9$\pm$2.4 \\
3 & reduced & S + F & (38)38/1097 & 3.5$\pm$0.6 & (37)31/446 & 7.0$\pm$1.2 & (25)20/253 & 8.0$\pm$1.8 & (28)20/193 & 10.4$\pm$2.3 \\ \hline
\hline
\end{tabular}
\end{center}
$^{a}$ See \S \ref{sec:method} for definitions of various samples
and TTV thresholds. 
The `reduced' sample is selected based on SNR, transit duration, 
and luminosity; the S, P, and F criteria are illustrated in Fig. \ref{fig:pttv}.
\\$^{b}$ Number of identified TTV candidates, versus numbers of KOIs
in that category.  The numbers in {\gr parentheses}
are the raw TTV candidates, while the  {\gr the corrected ones after parentheses result from}   
removing one of the two TTV candidates from {\gr the} count whenever a TTV pair is seen.
\\$^{c}$ {\w The measured} TTV fraction using the {\w corrected} TTV count.
\end{table*}

 
\subsection{ TTV Fraction \& Multiplicity }
\label{sec:ttv_frac}

We observe a remarkable rise of the TTV fraction
with transit multiplicity.

\begin{figure*}
\begin{center}
\includegraphics[width=0.9\textwidth, trim=20 30 20 20,clip]{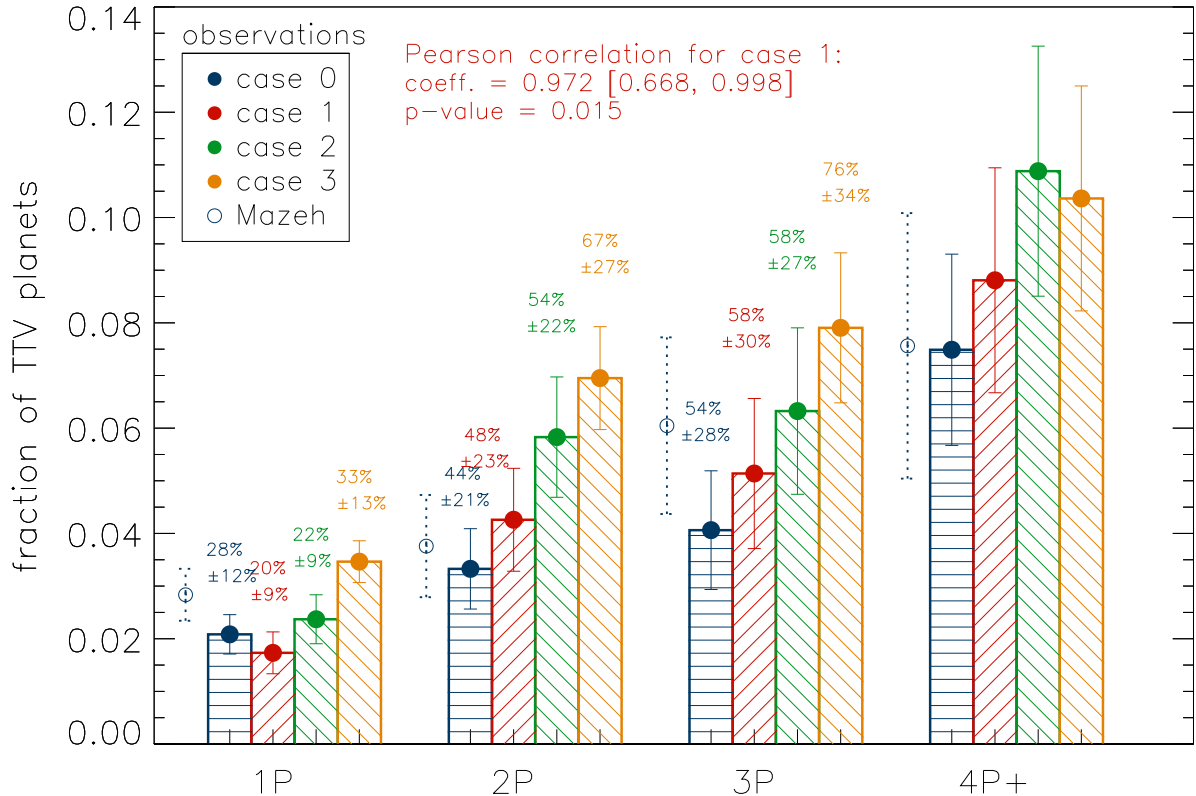}
\caption{The {\w measured} TTV fraction
  in 1P, 2P, 3P and 4P+ systems, for the 4 combination
    of selection thresholds described in Table 1.  The
  error bars are $1-\sigma$ Poisson error  in each case.  Within each
  transit group, the TTV fractions rise as we relax the criterion for
  identifying TTV candidates (from case 1 to 2 to 3). The TTV
  fractions for case 0 (full sample) are somewhat lower than for case
  1 (reduced sample), as we are more sensitive to the presence of TTV
  in the reduced sample. 
%
  The percentages printed above each column stand for the relative TTV
  fraction, normalized by that in the 4P+ group.  The occurrence rate
  of sinusoidal TTV is much lower in singles than those in higher
  multiples. Results obtained using the TTV catalogue of \citet{Maz13}
  {\wyq, with selection criteria equivalent to our case 0,} are shown
  as open circles {\wyq and} 
are consistent  with our values.  The Pearson statistics \citep{RN88} 
shows that
    the TTV fraction is positively correlated with the transit
    multiplicity (correlation $r=0.97$ with $1$ being perfect linear
    correlation), and that the correlation is  {\wyq
      statistically} significant
{\wyq    -- the null-hypothesis (no correlation) is strongly rejected with
a p-value of  $0.015$.} {\wyq From Monte-Carlo simulations ,
we obtain a $95\%$ confidence interval 
for the correlation coefficient from 0.668 to 0.998 given the error bars of  TTV fractions in case 1.}
}

\label{fig:frac}
\end{center}
\end{figure*}

In Table 1, we list the number of TTV candidates for different
transiting multiplicities, for four different combinations of sample
and TTV selection criteria.  For ease of comparison against theory, we
list the {\gr ``measured''} TTV fractions, obtained by
removing one candidate from {\gr the raw}
count whenever {\w both it and its TTV}
{\gr partner\footnote{ {\gr TTV partners are} two {\gr planets }
    that are near-MMR and share the same TTV super-period.} have observed TTVs.}  {\w Such
    a method is justified in \S \ref{sec:discussion}.} From now on, we
  focus on these {\w ``measured''} fractions (illustrated
  in Fig. \ref{fig:frac}) -- in fact, we focus on the relative  {\w ``measured''} fractions, the TTV fractions
  normalized by that in 4P+ systems. {\w The choice for the
    normalization is arbitrary. However, since the errorbars for these
    relative fractions are taken to be quadratic sums of the
    individual errorbars, which type of system one normalizes against
    does not affect the statistical conclusion.} These {\w results}
  are presented in Fig. \ref{fig:frac}.

  Except for case 3, all other combinations give very similar results
  for the relative TTV fractions: 1P systems have about {\gr five}
  times lower {\w (values from case 1: $20\pm 9\%$)} TTV
  fraction than 4P+, and 2P and 3P systems are about twice lower {\w
    ($48\pm 23\%$, and $58\pm 30\%$)}.  Results from case 3 are less
  reliable as the TTV selection criterion is too relaxed and allows
  for too many false positives.

We have also used the TTV data from \citet{Maz13},
published while we are editing our final draft, to confirm the above
results (Fig. \ref{fig:frac}).\footnote{\citet{Maz13} published a TTV
  catalog for 1897 KOIs.  They have also provided FAP values for TTV
  based on the LS periodogram. In Fig. \ref{fig:frac}, we plot TTV
  fractions from their catalog (their Table 3) after applying our `S'
  selection criterion. The good agreement is encouraging, as we
  extract TTVs using a different method.}

\begin{figure*}
\begin{center}
\includegraphics[width=0.99\textwidth]{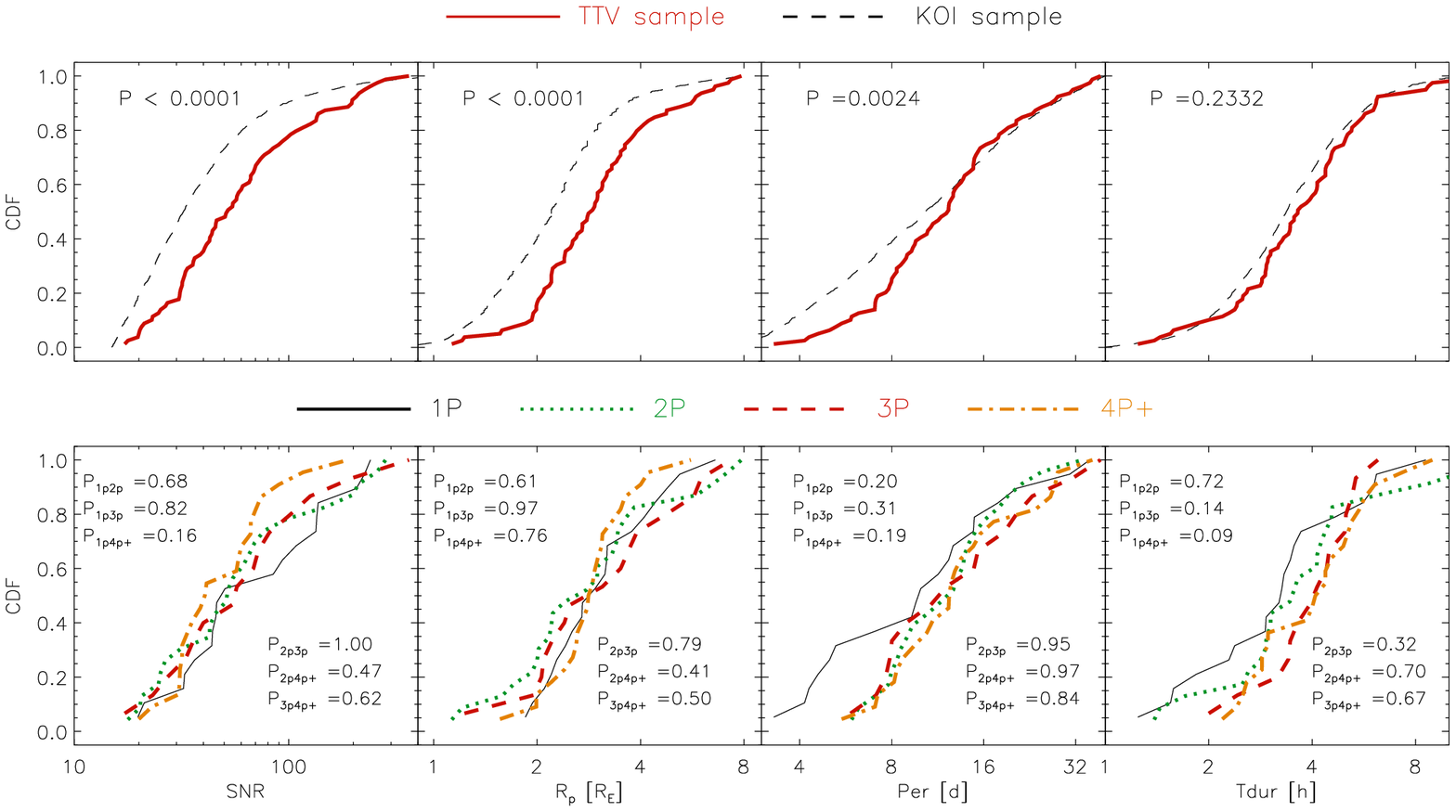}
\caption{Comparison of properties. The top panels compare those with
  sinusoidal TTV (red solid curve) to all KOIs (black dashed curve). 
The lower panels compare
  TTV candidates with different transit multiplicities.
  Here we only display comparison in transit signal-to-noise ratio
  (SNR), planet radius ($\rm R_{p}$), orbital period (Per) and {\w transit duration}, but we
  have performed a range {\w of} other ones (see text).  {\w TTV detection disfavours planets with small SNR and
    small radius ($R_p \leq 2 R_E$), and favours planets with
    intermediate orbital periods ($P \sim 10$ day).}  {\w In
    contrast,} TTV groups with different multiplicities {\w are statistically indistinguishable
    in} these {\w transit parameters
    (p-values from KS tests are listed)}.  
    }
\label{fig_dis_ttv}
   \end{center}
\end{figure*}

\subsection{Potential Bias}
\label{subsec:bias}

We first discuss what potential bias may affect the absolute TTV
fractions that we obtain, then move on to discuss biases that may
affect the relative TTV fractions among {\w groups of} different
transit multiplicity.  It becomes clear that by focusing only on the
relative TTV fractions, we can eliminate many {\w , if not all,}
observational bias.

We compare properties of the set of TTV candidates against the KOI
sample.  The top panels of Fig. \ref{fig_dis_ttv} display four {\w transit} properties: signal-to-noise ratio, planet
radius, orbital period and {\w transit duration}.
{\xie The TTV sample have in general larger transit SNR, larger planet
  radii} {\w (disfavouring small planets)} {\xie and slightly longer
  transit durations than the average KOIs.}  In addition, they are
also concentrated around orbital periods $\sim 10$ days. These
characteristics allow for optimum TTV detections {\w , as is
  demonstrated in recent TTV studies by \citet{For12b, Maz13}.}  For
instance, longer orbital periods generally lead to larger TTV
amplitudes {\xie \citep{HM05, Ago05,Lit12, Maz13}}, yet too long
orbital periods permit only a small number of transits to be
observed. As such, we expect that the {\w intrinsic} TTV
fraction, quantified as {\gr half} the fraction of planets that have
comparable TTV amplitudes as the ones detected here, should be higher
than our reported values (Table 1). 

On the other hand, we find no significant difference between the TTV
and KOI samples in terms of stellar mass, effective temperature,
metallicity and stellar brightness {\w \citep[stellar parameters
  from][]{Bat13}. KS tests performed to compare these two populations
  always return p-values greater than $0.5$.} This suggests that TTV
candidates live in all possible systems. {\w However, this deserves
  further study as currently there are large uncertainties in stellar
  parameters, and our TTV sample is relatively small.}

While the TTV sample as a whole are a biased representation of the KOI
sample, we find that the different sub-samples, separated by their transiting
multiplicity, {\w share similar
  distributions}
{\w in both the transit parameters (Fig.\ref{fig_dis_ttv}) and the
  stellar parameters (not shown here). The large p-values returned
  from KS tests (Fig.\ref{fig_dis_ttv}) do not support the hypothesis
  that the different subgroups experience different selection
  effects. }  {\w Moreover, we have confirmed that our reduced KOI
  samples, when separated into groups of different transit} {\w
  multiplicities, are statistically similar in their transit and
  stellar parameters}.  {\w Since the ability to detect TTV above a
  certain threshold amplitude, only depends on these transit and
  stellar parameters, these two results then argue that the relative
  measured TTV fractions reflect the relative intrinsic TTV fractions.}
{\w In other words, the significant correlation between TTV fraction
  and transit multiplicity that we observe (Fig.3) is unlikely to be
  caused by systematic biases  \xx{on stellar/transit parameters}. }
  
{\xx{ Another potential bias could {\wyq arise during}
    transit detection {\wyq --} transiting planets with
    significant TTVs can be systematically missed, or cataloged as
    false positives by the Kepler pipeline {\wyq \citet{GL11}}. To
    remove {\wyq this} bias,  \citet{CA13} {\wyq designed an algorithm (QATS)
      that} can simultaneously detect transits and measure their
    TTVs.  }
    {\xxx{Searches using QATS have only found a handful of  new planetary candidates (private communication J. A. Carter), which might indicate that the Kepler catalogue is not significant impacted by this bias. Nevertheless, we caution that there could be another possibility, namely, the QATS could not fully remove the bias, which deserves further study but is out of the scope of this paper. }}
 
{\xx{Last but not least, we note that the transit multiplicity of a
    given system is evolving as the catalog updates. For example, a 1P
    system {\wyq may} become a 2P system {\wyq when a new} transit candidate {\wyq is} found {\wyq ,
      either} due to accumulation of new data and/or improvement of
    the pipeline/algorithm for planet detection. To see how these
    factors affect our results{\wyq,} we took an older version KOI
    catalog from \citet{Bat13} and performed the same analysis as done
    in the standard case (case 1 in table 1). We obtained similar
    results for the relative TTV fractions: ($2.4\pm0.5)\%$,
    ($4.9\pm1.1)\%$, ($6.0\pm1.7)\%$ and ($10.3\pm3.0)\%$ for {\wyq
      the} 1P, 2P, 3P and 4P systems, respectively.  {\wyq This suggests} that the correlation between TTV
    fraction and transit multiplicity observed in Fig. 3
{\wyq may remain unaffected as more improved catalogues are published.} }}}

\section{Conclusion}
\label{sec:discussion}

\begin{figure}
\centerline{\includegraphics[width=0.45\textwidth]{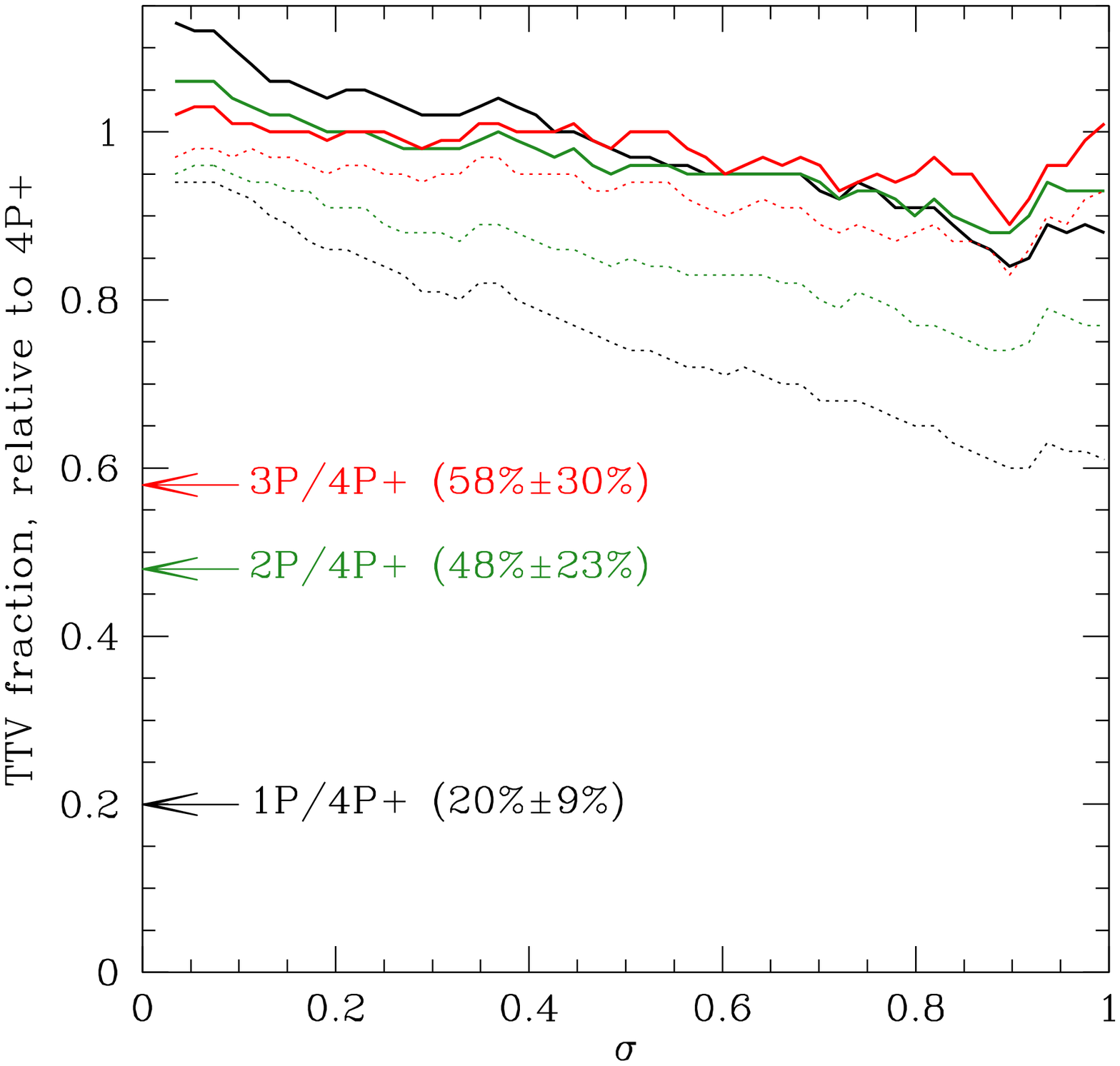}}
\caption{
  {\wyq Theoretical TTV fractions for different multiplicity groups,
    measured relative to that in 4P+ systems. Here, we construct a
    ensemble of planetary systems where the period ratio between
    neighbouring planets $x=P_2/P_1$, satisfies a Rayleigh
    distribution $ P(x) =x/\sigma^2\, e^{-x^2/2\sigma^2} $, and where
    the planets' mutual inclinations are described by an independent
    Rayleigh distribution with $\sigma_{\rm inc} = 3 \deg$. The
    results presented here are insensitive to the value of
    $\sigma_{\rm inc}$.  We identify TTV planets as those that transit
    their host stars and have a companion within a fractional distance
    of $2\%$ from a first-order MMR.  The solid lines indicate the
    measured fractions, while the dotted lines are the raw fractions
    (i.e., the fraction before removing doubly-counted TTV pairs),
    plotted as functions of $\sigma$, for different multiplicity
    groups.  The raw fractions in the 1P systems can drop to half of
    that in the 4P systems, because one is likely to observe both
    planets in the same TTV pair in the latter case. In contrast, in
    the corrected form (our so-called 'measured' TTV fraction), the
    relative TTV fractions remain close to unity, and are largely
    independent of either transit multiplicity or model parameters. In
    contrast, the observed relative fractions (case 1 in Fig.
    \ref{fig:frac}, marked here as arrows) fall much below unity.}  }
\label{fig:ttv_toymodel}
\end{figure}

We discuss the significance of our results by contrasting them against
predictions {\w from a simple toy model}.  {\w Assume} all KOIs, independent of transit
multiplicity, are drawn from the same intrinsic distribution, with
similar dispersions in mutual inclinations {\w and} planet spacing {\w (with no
  preference for MMRs)}. In this case, single systems are the ones
where the viewing angles are less favourable and we miss most of the
planets in the system, while the higher multiples are ones where more
planets are caught.  One can {\gr estimate} the TTV fraction for the theoretical
population {\gr as half} {\w the fraction of planets that both transit and
  have companions within a certain distance from a first-order MMR.}
{\w As} one naively expects and as is confirmed by {\gr Monte Carlo simulations (Fig. \ref{fig:ttv_toymodel})},
 {\w the relative TTV fractions} {\w cluster around $1$, and are largely independent of the
  model parameters and transit multiplicities.} 

The {\w observed} sharp rise of TTV fraction with {\w
  transit} multiplicity is inconsistent with such a simple
toy-model. {\w What are the possible interpretations?}

The lower TTV fraction observed for singles is unlikely to be
completely explained by the higher false positive rates in KOI
singles. The reported false positive rate is of order $10-20\%$
\citep{Fre13}. More importantly, our reduced sample, which is expected
to have a lower false positive rate than the full sample, yields the
same relative TTV fractions.  Moreover, since TTV amplitudes are
strongly boosted by eccentricities as small as a few percent {\xie
  \citep{HM05, Ago05, Ver11, Lit12}}, the lower TTV fraction can be
explained if higher multiple systems have higher
eccentricities. However, this is likely excluded by the tight spacing
observed among high multiples.

A simple explanation for our results is that {\xie the basic
  assumption} {\w in} {\xie our toy model is not true, namely,
  all KOIs cannot be treated as the same intrinsic population
  \citep{Lis11b, TD12, Joh12, Wei12}. For example, } 
  there could be at least two distinct
populations of Kepler planets, different in their intrinsic
frequencies of close companions.  The high multiples (4P+) are
dominated by a population that has a higher companion frequency, while
the 1P systems may be dominated by a population that have a lower
frequency of close companions. In other words, there are at least two
populations of Kepler planets, one that are closely spaced, and one
that is sparsely spaced.
  In an upcoming publication, we will use TTV fractions obtained in
  this paper, together with a variety of other observational facts, to
  constrain the properties of these two populations of Kepler
  planets. This will yield important {\gr constraints
    on} the process of planet formation.

\acknowledgements

JWX and YW acknowledge support by NSERC and the Ontario government.
Y.L. acknowledges support by NSF grant AST-1109776 and NASA grant NNX14AD21G

\bibliographystyle{apj}

\end{document}